\documentstyle[12pt]{article}
\input epsf
\begin{document}

\bibliographystyle{apsrev}
\def\half{{1\over 2}}
\def \D {\mbox{D}}
\def\curl {\mbox{curl}\,}
\def \ep {\varepsilon}
\def \lleq {\lower0.9ex\hbox{ $\buildrel < \over \sim$} ~}
\def \ggeq {\lower0.9ex\hbox{ $\buildrel > \over \sim$} ~}
\def\be{\begin{equation}}
\def\ee{\end{equation}}
\def\ber{\begin{eqnarray}}
\def\eer{\end{eqnarray}}
\def \apl {ApJ, }
\def \aps {ApJS, }
\def \pd {Phys. Rev. D, }
\def \prl {Phys. Rev. Lett., }
\def \pl {Phys. Lett., }
\def \np {Nucl. Phys., }
\def \l {\Lambda}

\def\apj{{Astroph.\@ J.\ }}
\def\mn{{Mon.\@ Not.\@ Roy.\@ Ast.\@ Soc.\ }}
\def\asta{{Astron.\@ Astrophys.\ }}
\def\aj{{Astron.\@ J.\ }}
\def\prl{{Phys.\@ Rev.\@ Lett.\ }}
\def\pd{{Phys.\@ Rev.\@ D\ }}
\def\nucp{{Nucl.\@ Phys.\ }}
\def\nat{{Nature\ }}
\def\plb {{Phys.\@ Lett.\@ B\ }}
\def \jetpl {JETP Lett.\ }
%........................................................

\begin{center}
\large{{\bf Modeling Repulsive Gravity with Creation}}
\end{center}

\begin{center}
R. G. Vishwakarma,\footnote{E-mail: rvishwa@mate.reduaz.mx}\\
\emph{Unidad Acad$\acute{e}$mica de Matem$\acute{a}$ticas\\
 Universidad Aut$\acute{o}$noma de Zacatecas\\
 C.P. 98068, Zacatecas, ZAC.\\
 Mexico}\\

\medskip

J. V. Narlikar\footnote{E-mail: jvn@iucaa.ernet.in}\\
\emph{Inter-University Centre for Astronomy \& Astrophysics\\
Post Bag 4, Ganeshkhind, Pune 411 007, India}\\

\end{center}

\medskip
\begin{abstract}
 There is a growing interest in the cosmologists for theories with
negative energy scalar fields and creation, in order to model a repulsive 
gravity. The classical steady state cosmology proposed by Bondi, Gold and 
Hoyle in 1948,
was the first such theory which used a negative kinetic energy 
creation field to invoke creation of matter.
We emphasize that creation plays very crucial role in cosmology and provides
a natural explanation to the various explosive phenomena occurring in local 
($z<0.1$) and extra galactic universe. We exemplify
this point of view by considering the resurrected version of this theory - the quasi-steady state theory, which  tries to relate creation events directly  
to the  large  scale  dynamics of the universe and supplies more natural explanations of the observed phenomena.  

Although, the theory predicts a decelerating universe at the present era,
it explains successfully the recent SNe Ia observations (which require an
accelerating universe in the standard cosmology), as we show in this paper by 
performing a Bayesian analysis of the data.
 
\medskip
\noindent
{\it Subject heading:} cosmology: theory, creation, negative energy fields, SNe Ia.

\noindent
{\bf Key words:} cosmology: theory, observation, creation, negative energy fields, SNe Ia.

\noindent
PACS: 98.80.-k,~98.80.Es,~98.90.+s
\end{abstract}

%\maketitle

\noindent
{\bf 1. INTRODUCTION}

\noindent
Remarkable progress has been made in various types of astrophysical
and cosmological observations in recent years. Among these, the accurate 
measurements of the anisotropies in the CMB made by the WMAP experiment appear
to offer the most promising determination of the cosmological parameters. 
The results of the WMAP experiment are however often quoted
as providing a direct evidence for an accelerating universe, which is though 
not correct. The cosmological constraints as established by the WMAP team  
(Spergel et al. 2003) entirely rely on the power law spectrum assumption and could be 
erroneous (Kinney 2001; Hannestad 2001). Taken on their face value, the WMAP observations
are fully consistent with the decelerating models like the CDM Einstein-de 
Sitter model (Vishwakarma 2003; Blanchard 2005).

 The possibility of an accelerating universe in fact emerges from the 
measurements of distant SNe Ia, which look fainter than they are 
expected in the standard decelerating models.
This observed faintness is generally explained by invoking some hypothetical 
source with negative pressure generally known as \emph{`dark energy'}.
This happens because the metric distance of an object, out to any redshift, 
can be increased by incorporating a \emph{`fluid'} with negative
pressure in Einstein's equations and hence the object looks fainter.
The simplest and the most favoured candidate of dark energy is a positive 
cosmological constant $\Lambda$, which is though plagued with the horrible 
fine tuning problems - an issue amply discussed in the literature.
This has led a number of cosmologists to resort 
to scalar field models called quintessence 
whose function is to cause the scale factor to accelerate at late times 
by violating the strong energy condition. 
While the scalar field models enjoy considerable popularity, they have not 
helped us to understand the nature of dark energy at a deeper level. 
By and large, the scalar field potentials used in
the literature have no natural field theoretical justification and have to be 
interpreted as a low energy effective potential in an ad-hoc manner.
Moreover they also require fine tuning of the parameters in order to be viable
(to find several other shortcomings, see for example, Padmanabhan 2005).

As desperate times call for desperate measures, the cosmologists, in order 
to model the dark energy, have now turned to \emph{`phantom'} or 
\emph{`ghost'} scalar field models with negative kinetic energy 
(Caldwell 2002; Carroll, Hoffman \& Trodden  2003; Gibbons 2003; Singh, Sami \& Dadhich 2003; Sami \& Toporensky 2004).
The classical steady state cosmology proposed by Bondi, Gold and Hoyle in 1948,
was the first such theory which used a negative kinetic energy creation field 
to invoke creation of matter. It is interesting to note that, distinct from 
all the existing big bang models at that time, this model predicted an 
accelerating universe. However, 
it is unfortunate that the theory was not given any credit 
(which it deserved, despite the difficulties associated with it) when 
the SNe Ia observations started claiming an accelerating universe
in 1998.

Once the cosmologists have lost their inhibitions about negative energy fields,
the time is ripe for considering the idea that the creation of matter plays 
important role in cosmology. We exemplify this point of view by considering
the resurrected version of the classical steady state theory, namely the
quasi-steady state cosmology (QSSC) which has not been given proper attention 
as it deserves. This theory was proposed by Hoyle, Burbidge and Narlikar
in 1993 (1993; 1995), wherein the introduction of negative kinetic energy 
scalar field is not ad-hoc but is required to ensure that matter creation
does  not violate  the law of conservation of matter and energy. 
However, first we emphasize that the idea of creation of matter is already
present in general relativity, though hidden behind some simplifying 
assumptions.

With a suitable Lagrangian for the source terms, the Einstein field equations
can be written as
\begin{equation}
R_{ik} - \frac{1}{2} g_{ik} R  =- 8 \pi
G\bigg[T_{ik}^{(\rm matter)}+T_{ik}^{(\Lambda)}+T_{ik}^{(\phi)}+ ....\bigg],\label{eq:feq}
\end{equation}
where we have considered the speed of light $c=1$. The only constraint on the
source terms, which is imposed by this equation, is the conservation of right
hand side through the Bianchi identities:
$[R_{ij}-\frac{1}{2} R g_{ij}]^{;j}=0=\left[T_{ij}^{(\rm matter)}+T_{ij}^{(\Lambda)}+T_{ij}^{(\phi)}+ ...\right]^{;j}$, implying that only the sum of all the 
energy-momentum tensors is conserved, individually they are not. If we take
them conserved separately, as is in practice among the cosmologists, it can be 
done only through the additional assumption of no interaction (minimal 
coupling) between different 
source fields, which though seems ad-hoc and nothing more than a simplifying 
assumption. On the contrary, interaction is more natural and is a fundamental 
principle. Of course some ideal cases are consistent with the idea of
minimal coupling, for example, $T_{ik}^{(\rm matter)}$ with a constant 
$\Lambda$. However, imposing this assumption on non-trivial cases would
result in  losing some important information. For example, taken on the face
value, a time-dependent $\Lambda$, with matter, implies matter
creation and results in a Machian model (Vishwakarma 2002a). However, if one
makes an additional assumption of no interaction between
$T_{ik}^{(\rm matter)}$ and $\Lambda(t)$, these features are lost and $\Lambda(t)$
reduces to a constant. It is well known that even if one considers the 
Robertson-Walker spacetime (to avoid non-Machian Godel's solution of 
Einstein field equations), there still exists a non-Machian solution of 
Einstein field equations - 
 the de Sitter solution. Creation has many more attractive features. It has
been shown how the scalar creation field helps in resolving the problems of
singularity, flatness and horizon in cosmology (Narlikar \& Padmanabhan 1985).
Such a negative energy creation field is responsible for a non-singular bounce
from a high non-singular density state, as has been shown by Hoyle and Narlikar
(1964). This idea has been recently used by Steinhardt and Turok in their
oscillatory model (Steinhardt \& Turok 2002).
The quasi-steady state cosmology (QSSC) is also a Machian theory which
is derived from an action principle based on Mach's Principle, and
assumes that the inertia of matter owes its origin to other matter
in the universe. The stress-energy tensor for creation (corresponding to
$T_{ik}^{(\phi)}$ in equation (\ref{eq:feq})) is given by
\be
T^{\rm creation}_{ik}= -f\left(C_i C_k+\frac{1}{4}C^\ell C_\ell ~ g_{ik}\right)\label{eq:tc},
\ee
where $f$ is a positive coupling constant and the gradient $C_i\equiv \partial \phi/\partial x^i$ is the contribution from a trace-free zero rest mass scalar field $\phi$ of \emph{negative} energy and stresses. The $\Lambda$ in this 
theory (corresponding to a $T_{ik}^{(\Lambda)}\equiv -\Lambda g_{ik}/8\pi G$ of (\ref{eq:feq})) 
appears as
a constant of nature with its value 
$\approx -2 \times 10 ^{-56} {\rm ~cm}^{-2}$,
which falls within the normally expected region of
the magnitude of the cosmological constant.  Note however that
its sign is negative, which is a consequence of the
Machian origin of the cosmological constant. 
The theory does not face the cosmological constant problem mentioned earlier.
In fact, the $\Lambda$ in the QSSC does not represent
the energy density of the quantum fields, as this model does not experience
the energy scales of quantum gravity except within the local centres of 
creation.
 The theory offers a purely stellar-based interpretation of all observed nuclei
including the light ones (Burbidge et al. 1957; Burbidge \& Hoyle 1998).
In the following, we demonstrate 
in brief the main features of this cosmology and how it confronts the various
observations (for more details, one can consult (Sachs, Narlikar \& Hoyle 1996; Hoyle, Burbidge \& Narlikar 2000)).

The QSSC represents a cyclic universe with its
Robertson-Walker scale factor given by
\begin{equation}
S(t) = e^{t/P}  [{1~+~\eta {\rm ~cos} (2\pi \tau/Q)}]\label{eq:sf},
\end{equation}
where the time scales $P~\approx 10^{3}$ Gyr $\gg Q~\approx 40-50 $
Gyr are considerably greater than the Hubble time scale of $10-15$ Gyr
of the standard cosmology.  The function $\tau(t)$ is very nearly like the cosmic
time $t$, with significantly different behaviour for short duration near
the minima of the function $S(t)$. The parameter $\eta$ has modulus
less than unity, thus preventing the scale factor from reaching zero.
Typically, $\eta \sim 0.8-0.9$. Hence there is no spacetime singularity, nor a violation of the law of
conservation of matter and energy, as happens at the big bang epoch in
the standard cosmology. The model has cycles of expansion
and contraction (regulated respectively by the creation field and the negative
$\Lambda$) of comparatively shorter period ($Q$) superposed on a long
term ($P$) steady state-like expansion. Creation of matter, which occurs
 through explosive processes, is
also periodic, being confined to pockets of strong gravitational fields around
compact massive objects and the nuclei of existing galaxies.
Such processes take place whenever
the energy of the creation field quantum rises above a threshold energy, which
is equal to the restmass energy of the created Planck particle.

The model provides
a natural explanation to the various explosive phenomena occurring in local 
($z<0.1$) and extra galactic universe.
By the early 1960s it had become clear that very large energy
outbursts are taking place in the nuclei of galaxies.
In the decades since then it has been found that many active
nuclei are giving rise to x-rays, and relativistic jets,
detected in the most detail as high frequency radio waves.  A very
large fraction of all of the energy which is detected in the
compact sources is non-thermal in origin, and is likely to be
incoherent synchrotron radiation or Compton radiation.
In addition to this  we see several other explosive
phenomena in the universe, such as jets from radio sources, gamma
ray bursts, X-ray bursters, QSOs, etc.
Generally it is assumed that a black hole plays the lead role in
such an event by somehow converting a fraction of its huge
gravitational energy into large kinetic energy of the `burst'
kind. In actuality however, we do not see infalling matter that is the
signature of a black hole.  Rather we see outgoing matter and
radiation, which agrees very well with the idea of creation events formulated 
in the framework of the QSSC.

There are several free parameters in the model which are estimated from the
observations and provide a decelerating universe at the present cycle of 
expansion. It is then interesting
to see how the model explains the SNe Ia and other observations! This is 
shown in the following.

\bigskip
\noindent
{\bf 2. The High Redshift Supernovae Ia}

\noindent
It is generally accepted that metallic vapours are ejected 
from the SNe
explosions which are subsequently pushed out of the galaxy through pressure
of shock waves (Hoyle \& Wickramasinghe 1988).
Experiments have shown that metallic vapours on cooling, condense into
elongated whiskers of $\approx$ $0.5-1$ mm length and $\approx$$10^{-6}$ cm 
cross-sectional radius (Donn \& Sears 1963; Nabarro \& Jackson 1958). It can be shown that the 
 extinction from the whisker dust adds an extra magnitude $\delta m(z)$ to the apparent 
magnitude $m(z)$ (arising from the cosmological evolution) of the SN light emitted 
at the epoch of redshift $z$, which is given by
\begin{equation}
\delta m(z)=1.0857\times ~\kappa ~\rho_{\rm g0}\int_0^z
(1+z')^2\frac{{\rm d}z'}{H(z')}.\label{eq:deltam}
\end{equation}
where $\kappa$ is the mass absorption coefficient and $\rho_{\rm go}$ is the 
whisker grain density at the present epoch. 
The net apparent magnitude is then given by
\begin{equation}
 m^{\rm net}(z)=m(z) + \delta m(z).\label{eq:mnet}
\end{equation}
By taking account of this effect, it has been shown that this kind of dust 
extinguishes radiation travelling over long distances and decelerating models 
without any dark energy (for example, the Einstein-de Sitter model)
can also explain high redshift SNe Ia observations successfully 
(Vishwakarma 2002b; 2003; 2005). QSSC in fact resorts to this dust to
explain not only the SNe Ia observations but also CMB, as we shall see in
the following. It has been shown (Narlikar, Vishwakarma \& Burbidge 2002; Vishwakarma \& Narlikar 2005) that by taking account 
of this effect, QSSC explains successfully the SNe Ia 
data from Perlmutter et al. (1999) and also shows an acceptable fit to
the `gold sample' of 157 SNe Ia recently published by Riess et al. (2004)
which, in addition to having previously observed SNe, also includes
some newly discovered highest-redshift SNe Ia by the Hubble Space Telescope.
Though this sample is believed to have a `high-confidence' quality of the spectroscopic 
and photometric record for individual supernovae, we note that there are some SNe
(1997as, 1997bj, 2000eg, 2001iw, 2001iv) in this sample which do not seem to be
consistent with any of the models generally considered in the fitting and appear 
as general outliers (see the encircled SNe in Fig. 1 and Fig. 2).
By excluding these points, the fit to different models improves considerably.
For example, the $\chi^2$ value per degrees of freedom (dof) for the best-fitting
QSSC model reduces to 1.18 from the earlier $\chi^2$/dof=1.30 obtained from the full
sample of 157 points (Vishwakarma \& Narlikar 2005). The fit to the standard (flat $\Lambda$CDM)
cosmology improves tremendously from  $\chi^2$/dof=1.14 (from 157 points) to 
$\chi^2$/dof=0.99 (from 152 points). The details of the fit (in the case of the frequentist approach) can be found in (Vishwakarma \& Narlikar 2005).

\newpage

\begin{figure}[tbh!]
\centerline{{\epsfxsize=14cm {\epsfbox[50 250 550 550]{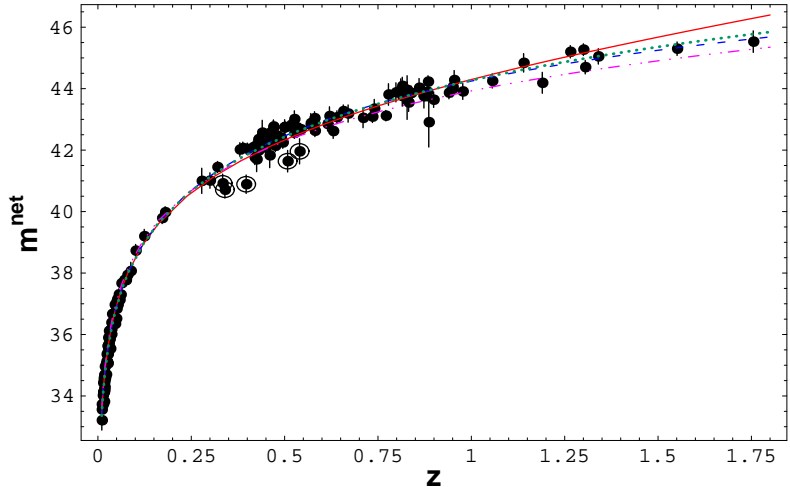}}}}
{\caption{\small Some best-fitting models are 
compared with the `gold sample' of SNe Ia data with 157 points as considered
by Riess et al. (2004). The solid curve corresponds to the QSSC model with the whisker dust, the 
dotted curve corresponds to the flat $\Lambda$CDM model, the dashed curve 
corresponds to the spherical $\Lambda$CDM model, and the dashed-dotted
curve corresponds to the Einstein-de Sitter model.
The models differ significantly for $z>1.2$. The encircled points seem to be general 
outliers which are missed by all the models.}}
 \end{figure}

\begin{figure}[tbh!]
\centerline{{\epsfxsize=14cm {\epsfbox[50 250 550 550]{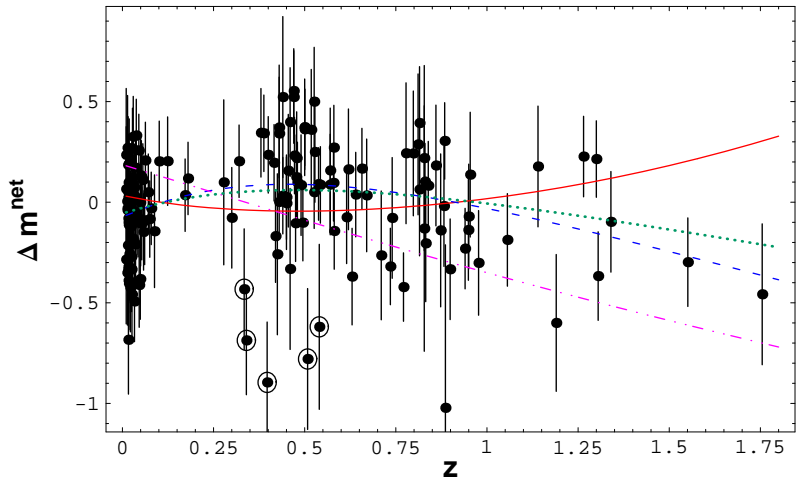}}}}
{\caption{\small Modified Hubble diagram of the `gold sample' of SNe Ia minus a 
fiducial model ($\Omega_{\rm m0}=0$, $\Omega_{\Lambda0}=0$). 
The relative magnitude($\Delta m^{\rm net}\equiv m^{\rm net}-
m_{\rm fiducial}$) is plotted for some best-fitting models, by using the original error bars.
The solid curve corresponds to the QSSC model with the whisker dust, the 
dotted curve corresponds to the flat $\Lambda$CDM model, the dashed curve 
corresponds to the spherical $\Lambda$CDM model, and the dashed-dotted
curve corresponds to the Einstein-de Sitter model. The encircled points seem to be general 
outliers which are missed by all the models. 
}}
 \end{figure}

Though there is not a clearly defined value of $\chi^2$/dof for an acceptable
fit, a \emph{`rule of thumb'} for a \emph{moderately} good fit is that $\chi^2$
should be roughly equal to the number of dof. A more quantitative measure
for the \emph{goodness-of-fit} is given by the $\chi^2$-\emph{probability}. 
If the fitted model
provides a typical value of $\chi^2$ as $x$ at $n$ dof, this probability is
given by
\be
Q(x, n)=\frac{1}{\Gamma (n/2)}\int_{x/2}^\infty e^{-u}u^{n/2-1} {\rm d}u.
\ee
Roughly speaking, it measures the probability that \emph{the model does
describe the data and any discrepancies are mere fluctuations which 
could have arisen by chance}. To be more precise, $Q(x, n)$ gives the probability 
that a model
which does fit the data at $n$ dof, would give a value of $\chi^2$ as large
or larger than $x$. If $Q$ is very small, the apparent discrepancies are
unlikely to be chance fluctuations and the model is ruled out. It may however
be noted that the $\chi^2$-probability strictly holds only when the models are
linear in their parameters and the measurement errors are normally distributed.
It is though common, and usually not too wrong, to assume that the 
$\chi^2$- distribution holds even for 
models which are not strictly linear in their parameters, and for this reason,
the models with a probability as low as $Q>0.001$ are usually deemed 
acceptable (Press et al. 1986). 
Models with vastly smaller values of $Q$, say, $10^{-18}$ are rejected.
The probability $Q$ for the best-fitting QSSC to the full sample is obtained as 0.007, 
which is though very small, but acceptable. By excluding the above-mentioned 5 outliers,
$Q$ improves to 0.062. The corresponding probabilities in the case of the standard
$\Lambda$CDM cosmology are obtained as 0.109 and 0.534.

We note that the fit to the QSSC is considerably worse than those in the standard
$\Lambda$CDM cosmology. However, one cannot compare the relative merits of the models
on the basis of the  $\chi^2$-probability (frequentist approach), which uses the 
best-fitting parameter values and hence judges only the maximum likely performance of 
the models. The more appropriate theory for such comparisons is the Bayesian theory
which does not hinge upon the  best-fitting parameter values and evaluates the overall 
performance of the models by using average likelihoods (rather than the maximum 
likelihoods), given by the \emph{Bayes factor} $B$. The theory employs the premise that if we assume an equal prior 
probability for
competing models, the probability for a given model is proportional to
the marginalised likelihood called {\it evidence}.
 We have described this theory,
in brief, in the Appendix (for more details, see (Drell, Loredo \& Wasserman 2000; John \& Narlikar 2002)).

In order to calculate the Bayes factor $B$ for the two models QSSC and the standard
$\Lambda$CDM, first we have to fix the prior probabilities for the free parameters.
While the flat QSSC has four free parameters $\kappa \rho_{\rm g0}H_0^{-1}$, 
$\Omega_{\Lambda0}$, $z_{\rm max}$ and ${\cal M}$ (see the Appendix of 
(Vishwakarma \& Narlikar 2005)),
the standard $\Lambda$CDM has only two free parameters $\Omega_{\rm m0}$ and 
${\cal M}$. We would like to mention that the whisker dust was
already introduced in the QSSC in order to explain the CMB which put a constraint
 on the density of the dust
$\rho_{\rm g0}\approx 10^{-34}$ g cm$^{-3}$ (Narlikar et al. 2003). This value, taken together
with the observational constraints on $\kappa$ (Wickramasinghe \& Wallis 1996) and $H_0$ (Freedman \& Turner 2003)
from other observations, supply a value of the parameter $\kappa \rho_{\rm g0}H_0^{-1}$
close to the best-fitting value estimated from the SNe Ia observations.
Hence we assign a prior on the parameter $\kappa \rho_{\rm g0}H_0^{-1}$ that it lies in 
the range $\kappa \rho_{\rm g0}H_0^{-1}\in [3.5, 6]$.
We also note that $\Omega_{\Lambda0}$ in the QSSC 
does not receive any significant contribution from  $\Omega_{\Lambda0}<-0.3$. Taking 
account of this and the theoretical constraint that $\Lambda$ in the QSSC is negative, 
we
assign a prior on the parameter $\Omega_{\Lambda0}\in [-0.3, 0]$. To the rest two 
parameters in the QSSC, about which we do not have prior information, we assign 
liberal priors: $z_{\rm max}\in[5, 10]$ (to be consistent with the highest redshift
$\approx 7$ observed so far) and ${\cal M}\in[41, 45]$ (which is the common parameter). 
For the parameter $\Omega_{\rm m0}$
in the flat $\Lambda$CDM model, we
assume that $\Omega_{\rm m0}\in$ [0, 1] (which is equivalent to assigning 
$\Omega_{\Lambda0}\in [0, 1]$). It may be noted that the $\Omega_{\Lambda0}$ in the 
two models are altogether different quantities (though they have been denoted by 
the same symbol in order to match the general convention) and there is no reason to
assign the same probability for them in the two different models.

When calculated for the full `gold sample' of 157 points,
these prior probabilities give a Bayes factor favouring
the standard $\Lambda$CDM over the QSSC as $B=2.77$, which though indicates an evidence 
against the QSSC, however, the evidence is not definite and is not worth more than 
a bare mention (for the interpretation of $B$, see the Appendix). 
Our assigning  $\Omega_{\Lambda0}\in [-0.3, 0]$ in the QSSC can raise eyebrows, as
this probability is very conservative compared with the one in the  $\Lambda$CDM.
However, assigning this probability is due to the reason that the likelihood for
$\Lambda$ in the QSSC 
does not receive any significant contribution from  $\Omega_{\Lambda0}<-0.3$, as mentioned earlier. For
example, increasing the domain of $\Lambda$ in its prior to 
$\Omega_{\Lambda0}\in [-1, 0]$
in the QSSC, results in lowering the likelihood of the model, as expected. This gives
a Bayes factor $B=9.30$, which indicates that the evidence against the QSSC is 
definite, though not strong. 

One should also note that a proper assessment of the probability for a model
is given by $p=1/(1+B)$. Thus for the above-mentioned two choices of the prior 
probabilities, the corresponding probabilities for the QSSC are 0.27 and 0.10,
which are reasonably good probabilities.

It is interesting to note that the whisker dust, which is a vital ingredient of the 
QSSC, does not make any significant improvement in the fit to the $\Lambda$CDM 
cosmology.
It may be argued that this kind of dust can create too much optical depth 
for the high redshift objects and they need to be excessively bright in
order to be seen. However, from our calculations, we find that the objects 
over the present cycle right upto the maximum redshift will be fainter, at the
most, by $\sim$6 magnitudes only and it is possible to see even sources from many 
previous cycles, though they will be very faint.

In Fig. 1, we have compared some best-fitting models with the actual data points.
In order to have a better visual comparison of these models, we
magnify their differences by plotting the relative magnitude with respect to
a fiducial model $\Omega_{\rm m0}=\Omega_{\Lambda0}=0$, without any 
whiskers (which has a reasonably good fit: $\chi^2$/dof $=1.2$). This has been shown
in the `modified' Hubble diagram in Fig. 2. In the following we describe, in brief,
the other important features of the QSSC.

\bigskip
\noindent
{\bf 3. The CMB}

\noindent
As far as the origin and nature of the CMB is concerned, the QSSC uses a
fact that is always ignored by standard cosmologists.  If we
suppose that most of the $^4$He found in our own and external
galaxies (about 24\% of the hydrogen by mass) was synthesized by
hydrogen burning in stars, the energy released amounts to about
4.37 x 10$^{-13}$ erg cm$^{-3}$.  This is almost exactly equal to
the energy density of the microwave background radiation with T =
2.74$^\circ$K! In the standard cosmology, this has to be dismissed
as a coincidence. 
Thus according to the QSSC, the CMB 
is the relic starlight left by the stars
of the previous cycles which has been thermalized by the metallic whisker 
dust emitted by the supernovae.
As a typical cycle proceeds from the maximum of scale factor towards the next minimum, 
the wavelengths of the starlight from the previous cycle are shortened since 
the universe contracts by a considerable factor. It has been shown
(Narlikar et al. 1997) that at wavelengths 100 m-20 cm, sufficient optical depth
exists for this radiation to thermalize in about twenty cycles.
 The production of microwaves in this
fashion goes on in each cycle, and the process of frequent absorption
and re-radiation by whiskers will eventually generate a uniform background,
{\it except for the contribution from the latest generation of clusters}.
These will stand out as inhomogeneities on the overall uniform background
arising from certain intrinsic inhomogeneities of the process as
well as from the cosmological model. A quantitative analysis shows that
this process requires an intergalactic dust of density of 
$\approx 10^{-34}$ g cm$^{-3}$, which is very close to the best-fitting value 
estimated from the SNe Ia observations.
The theory also explains  the peaks
at $l\sim 200$ and $l\sim 600$ which are related, in this cosmology, to the
clusters and groups of clusters (Narlikar et al. 2003).
Also, we have taken stock of the WMAP observations in (Narlikar, Burbidge \& Vishwakarma 2007). 

Though these studies do not give predictions as sharp as those given by the 
standard big bang cosmology, however, one should note the attitudinal 
difference between this approach and the standard one. In the big bang 
cosmology, the inferences are related to 
the postulated initial conditions prevailing
well beyond the range of direct observations (at redshifts 1100). 
Whereas the QSSC interpretation links the inhomogeneities of the radiation field to 
those of the matter field, 
on which we do not have very accurate data at present, but which may be 
observable one day.

It may also be worthwhile to mention that there are claims that like the dipole, the quadrupole 
and the octopole harmonics of the CMB spectrum also have their origin in the solar
system (Starkman et al. 2004). If this is correct, then subtracting this foreground
contribution from the rest of the signal (in order to have the temperature
fluctuations only at the time of the big bang) would render the inflationary
model in serious trouble.

\bigskip
\noindent
{\bf 4. The Non-Baryonic Dark Matter}

\noindent
Unlike the standard big bang cosmology, the QSSC allows the dark matter to be 
baryonic. It may be recalled that the standard cosmology predicts the 
existence of 
non-baryonic, though as yet undetected, particles to solve the problems of 
structure formation and of the missing mass in bound gravitational systems 
such as galaxies and clusters of galaxies.  
The most favoured candidate of non-baryonic dark matter postulated by many astrophysicists,
 cosmologists and particle physicists is a massive but very weakly interacting
 particle called WIMP (Weakly Interacting Massive Particle), a hypothetical 
elementary particle that was produced moments after the Big Bang.
Currently there are a number of WIMP detection experiments underway. Among these, the
DAMA experiment (Bernabei et al. 2003), which measures the annual modulation in WIMP interactions 
with the sodium-iodide detectors caused by the earth's rotation around the sun, is the 
only one to have claimed a positive signal. However, the results of this experiment are 
controversial as other more sensitive searches have not detected nuclear recoils due to 
WIMP interactions (Akerib et al. 2004; Angloher et al. 2005) and concluded that almost all the events measured by DAMA
were from neutrons, and should not be attributed to scattering events from dark-matter 
WIMPs. 
It is therefore fair to say that this scheme has still to demonstrate its viability.

However, in the framework of the QSSC, the dark matter need not be necessarily
non-baryonic. It can be in the form of baryonic matter being the relic of
very old stars of the previous cycles.
A typical QSSC cycle has a lifetime long enough for most
stars of masses exceeding $\sim 0.5-0.7 M_{\odot}$ to have burnt
out. Thus stars from previous cycles will be mostly extinct as
radiators of energy.  Their masses will continue, however, to
exert a gravitational influence on visible matter. The so-called
dark matter seen in the outer reaches of galaxies and within
clusters may very well be made up, at least in part, of these
stellar remnants.

It may  be timely to mention that the recent data on
distant x-ray clusters obtained from XMM and Chandra projects
indicate that the observed
abundances of clusters at high redshift, taken at face value, 
give $0.9<\Omega_{\rm m0}<1.07$ (at 1 $\sigma$) (Blanchard 2005). 
This favours a 
matter-dominated model and is consistent with the value of $\Omega_{\rm m0}$ 
in the QSSC estimated from the SNe Ia and CMB observations.
However, it is hard to reconcile with the concordance model.

\bigskip
\noindent
{\bf 5. Conclusion}

\noindent
In order to explain the current
observations in the framework of the standard cosmology, one has to
trust upon a preposterous composition for the constituent of the universe which
defies any simple explanation, thereby posing probably the greatest challenge 
theoretical physics has ever faced.
We think this is the right time to consider seriously
alternative theories  which present more natural explanations to
the observed phenomena, especially when there is no independent observational 
evidence for non-baryonic dark 
matter, dark energy and inflation, neither they have a firm basis in a well
established theory of particle physics.
Furthermore it is always necessary for healthy
science to have an alternative model to the dominant paradigm.

%\vspace{0.2cm}
\bigskip
\noindent
{\bf Acknowledgement} RGV thanks the Abdus Salam ICTP, for providing
hospitality and travel assistance (under his associateship programme), where
a part of the work was done.

\vspace{0.5cm}

\noindent
{\bf APPENDIX}

\setcounter{equation}{0}
\renewcommand{\theequation}{A.\arabic{equation}}

\noindent
The \emph{Bayes factor} is a  ratio of average likelihoods (rather
than the maximum likelihoods used for model comparison in frequentist 
statistics) for two models  $M_i$ and  $M_j$, and is given by
\be
B_{ij}=\frac{{\cal L}(M_i)}{{\cal L}(M_j)} \equiv\frac{p(D|M_i)}{p(D|M_j)},
\label{eq:BF}
\ee 
where the likelihood for the model $M_i$, ${\cal L}(M_i)$ is the probability
$p(D|M_i)$ to obtain the data $D$ if the model $M_i$ is the true one. For a model 
$M_i$ with free parameter, say, $\alpha$ and
$\beta$ (generalization for the models with more parameters is straight 
forward), this probability is given by 
\be
{\cal L}(M_i) \equiv p(D|M_i)=\int {\rm d}\alpha \int {\rm d}\beta~ 
p(\alpha|M_i) p(\beta|M_i) {\cal L}_i(\alpha, \beta),\label{eq:LP}
\ee 
where $p(\alpha|M_i)$ and $p(\beta|M_i)$ are the prior probabilities for the 
parameters $\alpha$ and $\beta$ respectively, 
assuming that the model $M_i$ is true. ${\cal L}_i(\alpha, \beta)$ is the 
likelihood for $\alpha$ and $\beta$  in the model $M_i$ and is given 
by the usual $\chi^2$-statistic:
\be
{\cal L}_i(\alpha, \beta)=\exp\left[-~\frac{\chi^2_i(\alpha, \beta)}{2}\right].
\ee 
For a flat prior 
probabilities for the parameters $\alpha$ and $\beta$, i.e., assuming that we 
have no prior information regarding $\alpha$ and $\beta$ except that they lie 
in some range [$\alpha$, $\alpha+\Delta\alpha$] and [$\beta$,
 $\beta+\Delta\beta$], we have $p(\alpha|M_i)=1/\Delta\alpha$ and 
$p(\beta|M_i)=1/\Delta\beta$. Hence the expression for the likelihood of the 
model $M_i$ reduces to
\be
{\cal L}(M_i) =\frac{1}{\Delta\alpha}\frac{1}{\Delta\beta}\int_\alpha^{\alpha+
\Delta\alpha} \int_\beta^{\beta+\Delta\beta} \exp\left[- ~\frac{\chi^2_i(
\alpha,\beta)}{2}\right]{\rm d}\beta ~ {\rm d}\alpha .\label{eq:likeli}
\ee
The \emph{Bayes factor} $B_{ij}$, given by (\ref{eq:BF}), which measures the 
relative merits of model $M_i$ over model $M_j$, is interpreted as follows
(Drell, Loredo \& Wasserman 2000; John \& Narlikar 2002; and the references therein). If
$1<B_{ij} <3$, there is an evidence against $M_j$ when compared with 
$M_i$, but it is not worth more than a bare mention.  If $3<B_{ij} <20$, the
evidence against $M_j$ is definite but not strong.  For $20<B_{ij} <150$, 
this evidence is strong and 
 for $B_{ij}>150$, it is very strong.

\begin{center}
{\bf References}
\end{center}

\noindent
Akerib, D.S., et al. 2004, Phys. Rev. Lett., {\bf 93}, 211301.\\               
Angloher, G., et al. 2005, Astropart. Phys. {\bf 23}, 325.\\
Bernabei, R., et al. 2003, Riv. Nuovo Cim., {\bf 26}, 1.\\ 
Blanchard, A. 2005, astro-ph/0502220.\\
Burbidge, E.M., Burbidge, G.R., Fowler, W.A., Hoyle, F. 1957, Rev. Mod. 

\hspace{.3cm} Phys., {\bf 29}, 547.

\noindent
Burbidge, G., Hoyle, F. 1998, ApJ, {\bf 509}, L1.\\
Caldwell, R.R. 2002, Phys. Lett. B, {\bf 545}, 23.\\
Carroll, S.M., Hoffman, M., Trodden, M. 2003, Phys. Rev. D, {\bf 68}, 023509.\\
Donn, B., Sears, G.W. 1963, Science, {\bf 140}, 1208.\\ 
Drell, P.S., Loredo, T.J., Wasserman, I. 2000, ApJ, {\bf 530}, 593.\\
Freedman, W.L., Turner, M.S. 2003, Rev. Mod. Phys., {\bf 75}, 1433.\\
Gibbons, G. 2003, hep-th/0302199.\\
Hannestad, S. 2001, Phys. Rev. D, {\bf 63}, 043009.\\
Hoyle, F., Burbidge, G., Narlikar, J.V. 1993, ApJ, {\bf 410}, 437.\\
Hoyle, F., Burbidge, G., Narlikar, J.V. 1995, Proc. Roy. Soc. A, {\bf 448},

\hspace{.3cm} 191.

\noindent
Hoyle, F., Burbidge, G., Narlikar, J.V. 2000, {\it A Different Approach to
 
\hspace{.3cm} Cosmology}, (Cambridge, Cambridge University Press); and the
 
\hspace{.3cm} references therein. 

\noindent
Hoyle, F., Narlikar, J.V. 1964, Proc. Roy. Soc. A, {\bf 278}, 465.\\
Hoyle, F., Wickramasinghe, N.C. 1988, Astrophys. Space Sc., {\bf 147}, 245.\\
John, M.V., Narlikar, J.V. 2002,  Phys. Rev. D, {\bf 65}, 043506.\\
Kinney, W.H. 2001, Phys. Rev. D, {\bf 63}, 043001.\\   
Nabarro, F.R.N., Jackson, P.J. 1958, in \emph{Growth and Perfection in Crystals}, 

\hspace{.3cm} eds. R. H. Duramus, et al, (J. Wiley, New York).

\noindent
Narlikar, J.V., Burbidge, G., Vishwakarma, R.G. 2007, `{\it Cosmology and Cos- 

\hspace{.3cm} mogony in a Cyclic Universe}', (submitted to JAA).

\noindent
Narlikar, J.V., Padmanabhan, T. 1985, Phys. Rev. D, {\bf 32}, 1928.\\
Narlikar, J.V., Vishwakarma, R.G., Burbidge, G. 2002, PASP, {\bf 114}, 1092 

\hspace{.3cm} (astro-ph/0205064).

\noindent
Narlikar, J.V., Vishwakarma, R.G., Hajian, A., Souradeep, T., Burbidge, G., 

\hspace{.3cm} Hoyle, F. 2003, Ap.J., {\bf 585}, 1 (astro-ph/0211036).

\noindent
Narlikar, J.V., Wickramasinghe, N.C., Sachs, R., Hoyle, F. 1997, Int. J. 

\hspace{.3cm} Mod. Phys. D, {\bf 6}, 125.

\noindent
Padmanabhan, T. 2005, in {\it 100 Years of Relativity - Spacetime Structure: 

\hspace{.3cm} Einstein and Beyond}, ed. A. Ashtekar, (World Scientific, Singapore), 

\hspace{.3cm} (gr-qc/0503107).

\noindent 
Perlmutter, S., et al. 1999, ApJ, {\bf 517}, 565.\\
Press, W.H., Teukolsky, S.A., Vetterling, W.T., Flannery, B.P. 1986, {\it Numer-

\hspace{.3cm} ical Recipes}, (Cambridge University Press). 

\noindent
Riess, A. G., et al. 2004, ApJ, {\bf 607}, 665.\\
Sachs, R., Narlikar, J.V., Hoyle, F. 1996,  A\&A, {\bf 313}, 703.\\
Sami, M., Toporensky, A. 2004, Mod. Phys. Lett. A, {\bf 19},  1509.\\
Singh, P., Sami M., Dadhich, N. 2003, Phys. Rev. D, {\bf 68}, 023522.\\ 
Spergel, D.N., et al. 2003, ApJS, {\bf 148}, 175.\\
Starkman, G., et al. 2004, Phys. Rev. Lett., {\bf 93}, 221301.\\ 
Steinhardt, P.J., Turok, N. 2002, Science, {\bf 296}, 1436.\\
Vishwakarma, R.G. 2003, MNRAS, {\bf 345}, 545 (astro-ph/0302357).\\
Vishwakarma, R.G. 2002a, Class. Quant. Grav., {\bf 19}, 4747 (gr-qc/0205075).\\
Vishwakarma, R.G. 2002b, MNRAS, {\bf 331}, 776 (astro-ph/0108118).\\
Vishwakarma, R.G. 2005, MNRAS, {\bf 361}, 1382 (astro-ph/0506217).\\
Vishwakarma, R.G., Narlikar, J.V. 2005, Int. J. Mod. Phys. D, {\bf 14}, 345 

\hspace{.3cm} (astro-ph/0412048).

\noindent
Wickramasinghe, N.C., Wallis, D.H. 1996, Astrophys. Space Sc.,
      {\bf 240}, 157.

\end{document}